\documentclass[12pt,preprint]{aastex}
\usepackage{graphicx}

\newcommand{\kms}{km s$^{-1}$}

\slugcomment{Submitted to \apj}
\usepackage{rotating}

\shorttitle{Star formation in NGC~3227 Tidal Dwarf Candidate}
\shortauthors{Mundell et al.}

\begin{document}


\title{The Unusual Tidal Dwarf Candidate in the Merger System
NGC~3227/6: Star Formation in a Tidal Shock?}


\author{Carole G. Mundell\altaffilmark{1}}
\affil{Astrophysics Research Institute, Liverpool John Moores
University, Twelve Quays House, Egerton Wharf, Birkenhead, CH41 1LD,
U.K.}
\email{cgm@astro.livjm.ac.uk}
\author{Phil A. James}
\affil{Astrophysics Research Institute, Liverpool John Moores
University, Twelve Quays House, Egerton Wharf, Birkenhead, CH41 1LD,
U.K.}
\email{paj@astro.livjm.ac.uk} 
\author{Nora Loiseau}
\affil{XMM-Newton Science Operations Center, VILSPA, Apartado 50727,
28080 Madrid, Spain} 
\email{nloiseau@xmm.vilspa.esa.es}
\author{Eva Schinnerer\altaffilmark{2}}
\affil{National Radio Astronomy Observatory, P.O. Box 0, Socorro, NM
87801-0387}
\email{eschinnerer@nrao.edu}
\and
\author{Duncan A. Forbes}
\affil{Centre for Astrophysics and Supercomputing, Swinburne
University of Technology, Mail Number 31, P.O. Box 218, Hawthorn, VIC 3122, Australia} 
\email{dforbes@astro.swin.edu.au}


\def\fsec{\hbox{$.\!\!^{s}$}}
\def\farcs{\hbox{$.\!\!^{\prime\prime}$}}
\def\hi{\hbox{H\,{\sc i}}}
\def\ha{\hbox{H$\alpha$}}
\def\H2{\hbox{H$_{\rm 2}$}}
\def\cmsq{cm$^{-2}$}
\def\nh{$N_{\mbox{\scriptsize H}}$}
\def\kms{km~s$^{-1}$}
\def\mJyb{mJy beam$^{-1}$}
\def\fdeg{\hbox{$.\!\!^{\circ}$}}
\def\fsec{\hbox{$.\!\!^{s}$}}
\def\arcmin{\hbox{$^\prime$}}
\def\arcdeg{\hbox{$^\circ$}}
\def\lesssim{\mathrel{\hbox{\rlap{\hbox{\lower4pt\hbox{$\sim$}}}\hbox{$<$}}}}
\def\moresim{\mathrel{\hbox{\rlap{\hbox{\lower4pt\hbox{$\sim$}}}\hbox{$>$}}}}

\altaffiltext{1}{Royal Society University Research Fellow}
\altaffiltext{2}{Jansky Postdoctoral Fellow at the National Radio
Astronomy Observatory}
 

\begin{abstract}
We report the discovery of active star formation in the \hi\ cloud
associated with the interacting Seyfert system NGC~3227/NGC~3226 that
was originally identified as a candidate tidal dwarf galaxy (TDG) by
Mundell et al. and that we name J1023+1952.  We present broad-band
optical B, R, I (from the INT) and ultraviolet images (from XMM-Newton)
that show the \hi\ cloud is associated with massive on-going star
formation seen as a cluster of blue knots (M$_B$$\lesssim$$-$15.5 mag)
surrounded by a diffuse ultraviolet halo and co-spatial with a ridge
of high neutral hydrogen column density
(\nh~$\sim$3.7~$\times$~10$^{21}$~\cmsq) in the southern half of the
cloud. We also detect \ha\ emission from the knots with a flux density of
F$_{H\alpha}$~$\sim$2.55~$\times$~10$^{-14}$~erg~s$^{-1}$~\cmsq\
corresponding to a star-formation rate of
SFR(H$_{\ha})$~$\sim$10.6~$\times$10$^{-3}$~M$_{\odot}$~yr$^{-1}$.
J1023+1952 lies at the base of the northern tidal tail, and,
although it spatially overlaps the edge of the disk of NGC~3227,
Mundell et al. showed that the \hi\ cloud is kinematically distinct
 with an \hi\ mean velocity 150 \kms\ higher than that of
 NGC~3227. Comparison of
 ionized (\ha) and neutral (\hi) gas kinematics of the cloud show
 closely matched recessional velocities, providing strong evidence
 that the star-forming knots are embedded in J1023+1952 and are not
 merely optical knots in the background disk of NGC~3227, thus
 confirming J1023+1952 as a gas-rich (M$_H$/L$_B$~$>$~1.5) dwarf
 galaxy.  No star formation is
 detected in the northern half of the
 cloud, despite similar \hi\ column densities; instead, our new high
 resolution \hi\ image shows a ridge of high column density
 coincident with
 the reddest structures evident in our B$-$I image. We suggest these
 structures are caused by the background
 stellar continuum from the disk of NGC~3227 being absorbed by dust
 intrinsic to J1023+1952, thus placing J1023+1952 in front of NGC~3227
 along the line of sight.  We discuss two scenarios for the origin of
 J1023+1952; as a third, pre-existing dwarf galaxy involved in
 the interaction with NGC~3227 and NGC~3226, or a newly-forming dwarf
 galaxy condensing out of the tidal debris removed from the gaseous
 disk of NGC~3227. The first scenario is feasible given that NGC~3227
 is the brightest member of a galaxy group, an environment in which
 pre-existing dwarf galaxies are expected to be common. However, the
 lack of a detectable old stellar population in J1023+1952 makes a
 tidal origin more likely.  If J1023+1952 is a bound object forming
 from returning gaseous tidal tail material, its unusual location at the base
 of the northern tail implies a dynamically young age similar to its
 star-formation age, and suggests it is in the earliest stages of TDG
 evolution.  Whatever the origin of J1023+1952 we suggest that its star formation is shock-triggered by collapsing tidal debris.

\end{abstract}




\section{INTRODUCTION}

The importance of collisions between galaxies as a key driver of
galaxy formation and evolution is well recognized, with the galaxy
merger rate being higher in the early Universe (Burkey et
al. 1994; Struck 1999) and galaxies undergoing several interactions,
of varying strengths, in their lifetime. Dwarf galaxies in particular
are thought to play an important role in galactic formation and
evolution but the origin of today's dwarfs galaxies is still debated;
cold dark matter hierarchical models invoke the merger of small dark
matter halos formed in the early Universe to produce larger galaxies
while alternative `down-sizing' models suggest large galaxies formed
first via collapse of primordial gas clouds and smaller galaxies were
created later, perhaps through tidal interaction (Cowie et al. 1996;
Hunter, Hunsberger \& Roye 2000).  Indeed, as many as one-half of the
current dwarf galaxy population of compact groups could have formed
from the interaction of giant spiral galaxies (Hunsberger, Charlton \&
Zaritsky 1996).

Tidal forces between interacting galaxies redistribute galactic
material, producing extensive stellar and gaseous tidal tails
(e.g. Toomre \& Toomre 1972; Barnes \& Hernquist 1992; Mihos 2001);
tidal dwarf galaxies (TDG) are small, gas-rich galaxies formed from
gas removed from the progenitor galaxy disks either as coherent gas
clouds ejected during the encounter (Elmegreen, Kaufman \& Thomasson
1993) or via the gravitational collapse of tidal debris (Zwicky 1956;
Duc et al. 2000).  As nearby galaxies in the process of formation,
TDGs offer an important laboratory to study the ongoing processes of
galaxy evolution and the link with tidal interaction, but to date only
a small number have been confirmed and studied in detail (e.g. Braine
et al. 2001).

NGC~3227 is a nearby Seyfert galaxy that is interacting with a
gas-poor dwarf elliptical companion NGC~3226. Neutral hydrogen (\hi)
imaging of the system (Mundell et al. 1995) revealed tidal tails
extending $\moresim$100 kpc north and south of the disk of NGC~3227
and well-ordered gas in the disk of NGC~3227; no \hi\ was detected in
NGC~3226. In addition to the tidal tails, Mundell et al. (1995)
discovered a 10$^8$~M$_{\odot}$ cloud of \hi\ close to, but physically
and kinematically distinct from, the galactic disk of NGC~3227. They
concluded this cloud was either a third galaxy in the interacting
system being accreted by NGC~3227 or that it represented gas stripped
from the disk of NGC~3227 and therefore was a candidate tidal dwarf
galaxy.

Here we present new ground-based broad-band BRI and narrow-band
H$\alpha$+[N{\sc ii}] optical imaging, ultraviolet continuum imaging
taken with the optical monitor on board XMM-Newton, and high angular
resolution $\lambda$21-cm (\hi) imaging of this candidate tidal dwarf
galaxy, named hereafter J1023+1952. We also present 2-D \ha\ ionized gas
kinematics of this object and compare them with the \hi\ kinematics.
We assume a distance to the NGC~3227 system of 20.4 Mpc (Tully 1988);
one arcsecond therefore corresponds to 99~pc.   

\section{OBSERVATIONS AND REDUCTION}
\label{observations}

\subsection{$\lambda$21-cm \hi\ Emission}

The $\lambda$21-cm neutral hydrogen emission from the NGC~3227 system
was imaged on 1995 November 26 and 30 using the Very Large Array (VLA)
in B configuration. The total on-source observing time was 10.5 hours
and data editing, calibration and analysis followed standard methods
(e.g. Mundell et al. 1999).  The final calibrated image cube has
1\farcs5 pixels, a circular restoring beam size of 6\farcs3 (624 pc),
a velocity resolution of 10.3 \kms\ and an r.m.s. noise per channel of
0.3 mJy~beam$^{-1}$.  Moment analysis (e.g. Mundell et al. 1999) was
performed on the image cube to isolate \hi\ emission from the cloud
J1023+1952, which is spatially and kinematically distinct from that of
the disk of NGC~3227, and to derive images of the integrated \hi\
intensity (Figure \ref{mom0}) and velocity structure of the cloud
(Section \ref{sectha}).

\subsection{Optical Continuum Imaging}

NGC~3227 was imaged on 2000 February 5 with the Wide Field Camera
(WFC) at the 2.5-m Isaac Newton telescope located on La Palma, Canary
Islands. The WFC has a pixel scale of 0\farcs33 per pixel.  We
obtained B, R and I images of 3$\times$1000~s, 2$\times$500~s and
3$\times$300~s exposure times respectively. The seeing was $\sim$1
arcsec. Due to partial cloud cover through the night the final photometry
was derived from values given for NGC~3227 in McAlary et
al. (1983). The data reduction was carried out using standard IRAF
tasks (e.g Miles et al. 2004, in prep). The B-band image of the
interacting system is shown in Figure \ref{HIonB}a; Figure
\ref{HIonB}b shows a close-up of the B-band image in the region of
J1023+1952 with \hi\ emission overlaid in contours.

\subsection{H$\alpha$ Imaging and Spectroscopy}


Narrow-band \ha\ observations of NGC~3227 were obtained using the
4.2-m William Herschel Telescope on La Palma. The TAURUS filter
centered at 6589 \AA, width 15 \AA, was used for a 600 s exposure;
R-band continuum data were used for the continuum subtraction (Loiseau
et al. in preparation). The pixel size is 0\farcs28.  The quoted flux
was the sum of the fluxes measured through four circular apertures
centered on the H$\alpha$-emitting regions co-spatial with J1023+1952
(Table \ref{props}) using the GAIA package.

H$\alpha$ spectroscopic images were obtained using the wide-field
Fabry-Perot interferometer
TAURUS-2\footnote{$(http://www.ing.iac.es/Astronomy/observing/manuals/html\_manuals/wht\_instr/taurus/taurus.html)$}
on the WHT (Loiseau et al. in preparation). After reduction and
wavelength calibration the resulting data cube had dimensions of 500 x
500 x 60 pixels, a pixel size of 0\farcs54/pixel and channel
separation of 14.1~\kms. The channel maps were combined to produce the
\ha\ velocity field (Section \ref{sectha}).

\subsection{Ultraviolet Continuum Imaging}

We obtained archival ultraviolet images taken on 2000 November 28
(P.I. Jansen) with the optical monitor on XMM-Newton in the UVM2
($\lambda$$_{c}$$\sim$2298~\AA, $\Delta\lambda$$\sim$439~\AA) and UVW1
($\lambda$$_{c}$$\sim$2905~\AA, $\Delta\lambda$$\sim$620~\AA) filters
(see Mason et al. 2001). The exposure time for each filter was 1~ks, the
pixel size is 0\farcs95 and the data were processed using SAS
version~5.4. Measured counts in each filter were converted to flux
units using a value averaged over different spectral types as listed
in Section 7.1.2.8 (Method 2 by Breeveld) in the SAS Watchout
documentation\footnote{(see
$http://xmm.vilspa.esa.es/external/xmm\_sw\_cal/sas.shtml$)}.

\subsection{Infrared Continuum Imaging}

Infrared images in J, H and K bands were obtained using the 1$-$5$\mu$m
imager-spectrometer, UIST, on the UK Infared Telescope (UKIRT) in
imaging mode on 2003 May 25. Exposures were taken with the 180\arcsec\
field of view (0\farcs12 pixels) centred on the location of J952+1026
and total integration times of 2220, 2220 and 3360 s in J, H and K
bands respectively.  Data reduction was performed using the ORAD-DR
pipeline. Infrared emission was detected from the edge of the disk of
NGC~3227 but no infrared emission was detected coincident with
J1952+1026.

\section{DUST, GAS AND STARS IN J1023+1952}

\subsection{$\lambda$21-cm \hi\ Emission}
\label{secthi}

\begin{figure} 
\epsscale{0.8}
\plotone{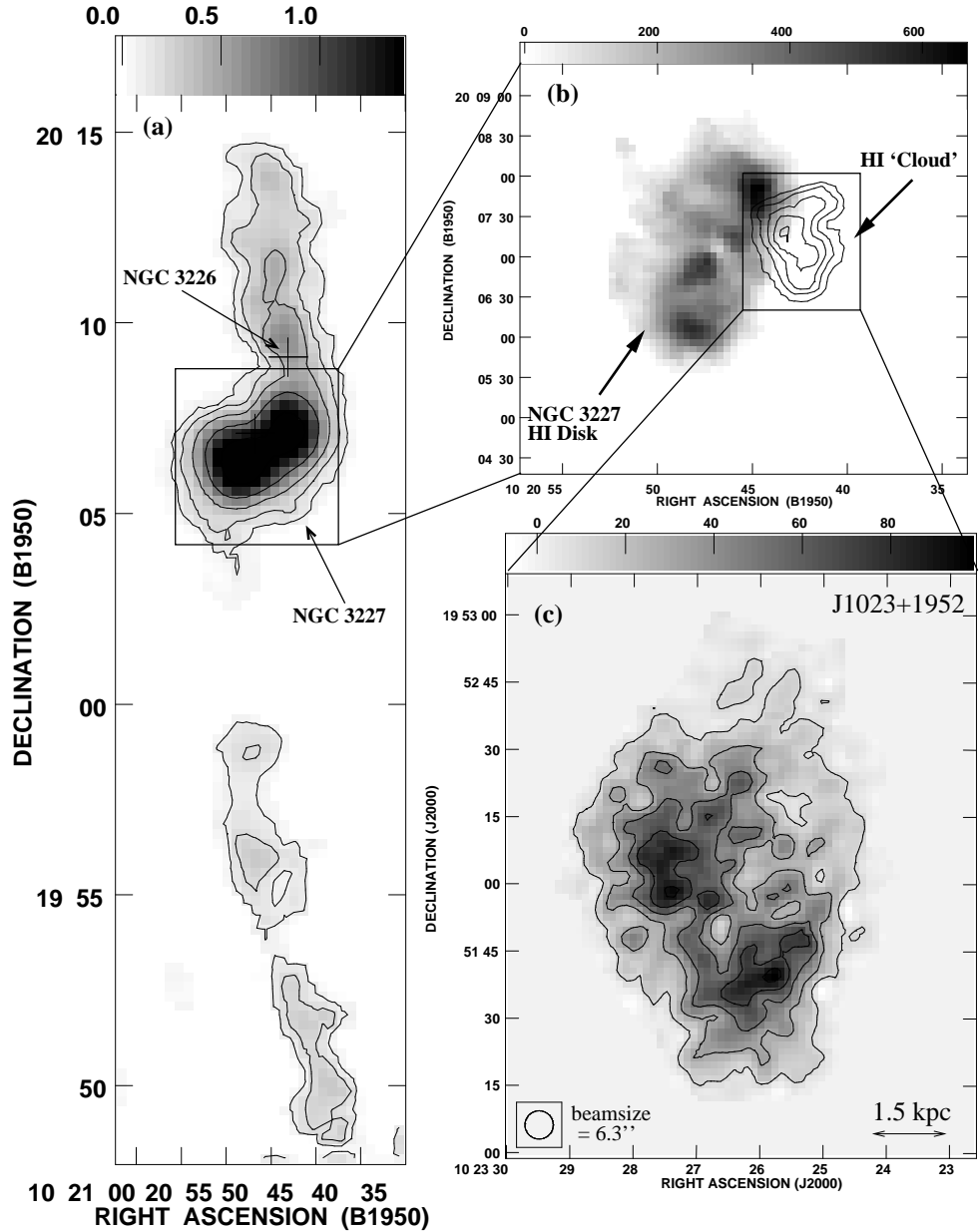} 
\caption{(a) VLA D-configuration HI image (zeroth moment) of the NGC~3227
system (from Mundell et al. 1995); the angular resolution is
60\farcs7~$\times$~58\farcs2 and the location of NGC~3226 is indicated
by  `$+$'; (b) VLA C-configuration HI imaging that resolved
HI emission from the disk of NGC~3227 and the 'HI cloud' (Mundell
et al. 1995); the angular resolution is 20\farcs5~$\times$~18\farcs3;
(c) new B-configuration image of integrated hydrogen emisson from the
dwarf candidate, J1023+1952. The angular resolution is
6\farcs3~$\times$~6\farcs3 and contour levels correspond to column
densities \nh\ = (0.4, 1.2, 2.1, 2.9, 3.7)$\times$10$^{21}$ \cmsq; peak \nh\ =
4.0$\times$10$^{21}$ \cmsq. 
\label{mom0}}
\end{figure}

Figure \ref{mom0}(c) shows the high angular resolution image of the
integrated \hi\ emission from J1023+1952; the distribution of \hi\ in
the NGC~3227 system at lower resolution, taken from Mundell et
al. (1995), is shown in Figures \ref{mom0}(a, b). J1023+1952 has an
\hi\ size of $\sim$90\arcsec~$\times$~60\arcsec\
(8.9~kpc~$\times$~5.9~kpc) and significant structure is evident within
the cloud with two prominent ridges of high \nh\ lying along
P.A.$\sim$0\arcdeg\ in the northeast and P.A. $\sim$120\arcdeg\ in the
south-west regions, hereafter: northern and southern ridges. More
diffuse emission extends to the northeast and connects to the northern
tidal tail reported by Mundell et al. (1995). Column densities in
J1023+1952 are in the range of $\sim$4~$\times$~10$^{20}$~$<$~\nh\
~$<$~4.0~$\times$~10$^{21}$~\cmsq\ (or 0.4~$<$~A$_B$~$<$~3.4 mag);
the northern and southern ridges have similar \nh\ with the peak \nh\
occurring in the southern ridge (Figure \ref{mom0}(c)).  \hi\ emission
is detected over the heliocentric velocity range 1188$-$1336 \kms\
with a full width at 20\% intensity of the integrated \hi\ profile of
W$_{20}$~$\sim$~120~\kms. These \hi\ properties are
consistent with those measured by Mundell et al. (1995) from the lower
resolution ($\sim$20\arcsec) C-configuration dataset; the integrated
\hi\ emission measured from our new \hi\ data is $\sim$12\% lower than
that measured by Mundell et al. (1995) - a discrepancy that may be due
to the inclusion of a small amount of faint extended emission from
tidal tail gas in the value quoted by Mundell et al. (1995). Since the
difference is small, we use the \hi\ mass taken from Mundell et
al. (1995) as an upper limit; correcting to a distance of 20.4~Mpc,
the \hi\ mass of J1023+1952 is therefore
M$_H$~$\sim$~3.8~$\times$~10$^8$~M$_{\odot}$.

\subsection{Optical Continuum Emission}

\begin{figure} 
\plotone{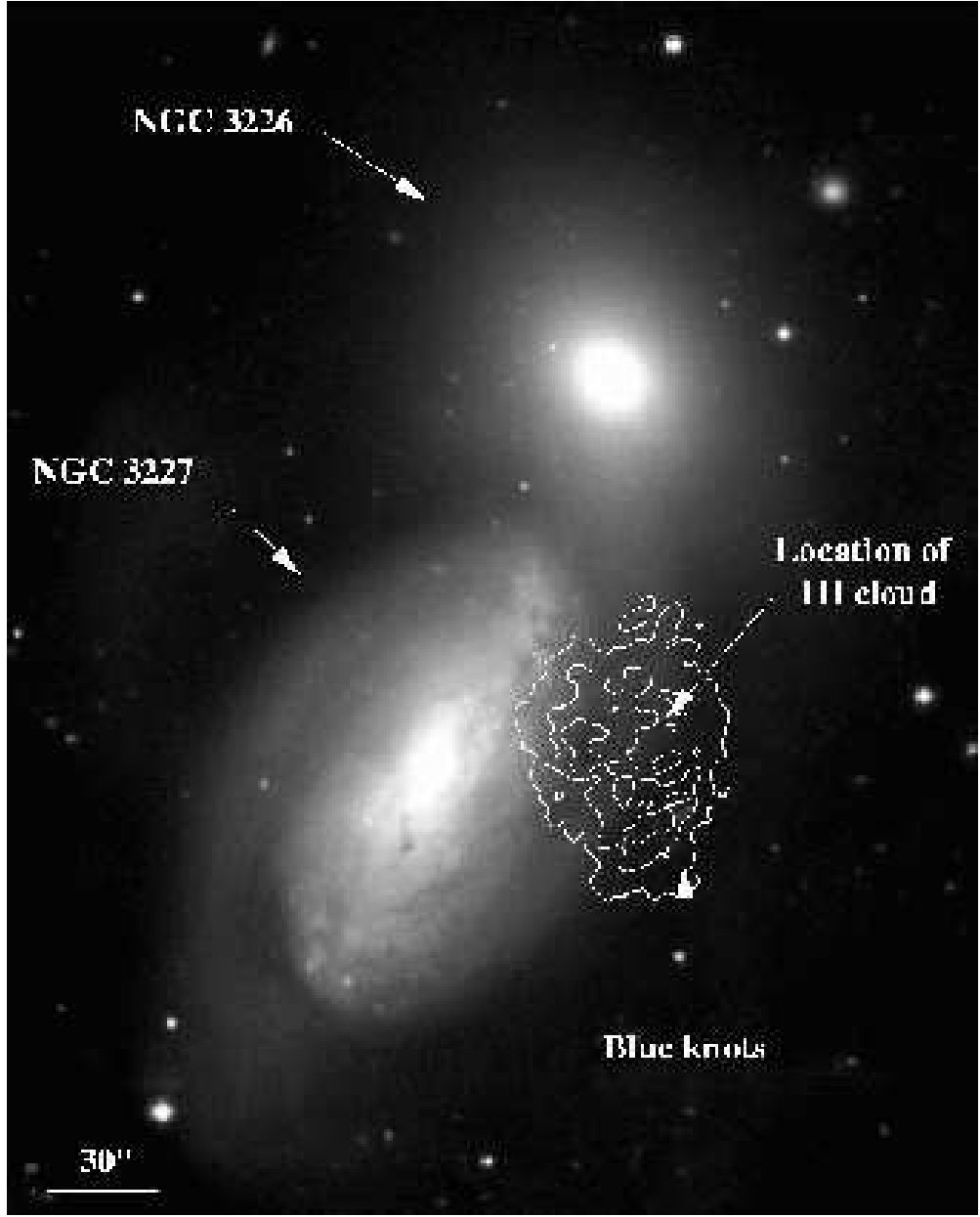}
\caption{Three-color image of the NGC~3227/3226 system. B,R, and I images are
in blue, green and red channels respectively. Contours of \hi\ emission from J1023+1952 are overlaid in white.
\label{threecol}}
\end{figure}

Figure \ref{threecol} shows the combined BRI three-color image of the
NGC~3227 system; continuum emission from NGC~3227 and NGC~3226
dominates the image and the two galaxies appear connected by a tidal
arm curving anticlockwise from the southern point of the NGC~3227 disk
to NGC~3226.  A string of blue optically-emitting knots can be seen
$\sim$70\arcsec\ ($\sim$7 kpc) west of the nucleus of NGC~3227; the
string extends over $\sim$27\arcsec\ (~$\sim$2.7 kpc) along
P.A. $\sim$120\arcdeg. Complex dust patches are also evident around
and to the north east of the string of knots.

\begin{sidewaysfigure}
\plotone{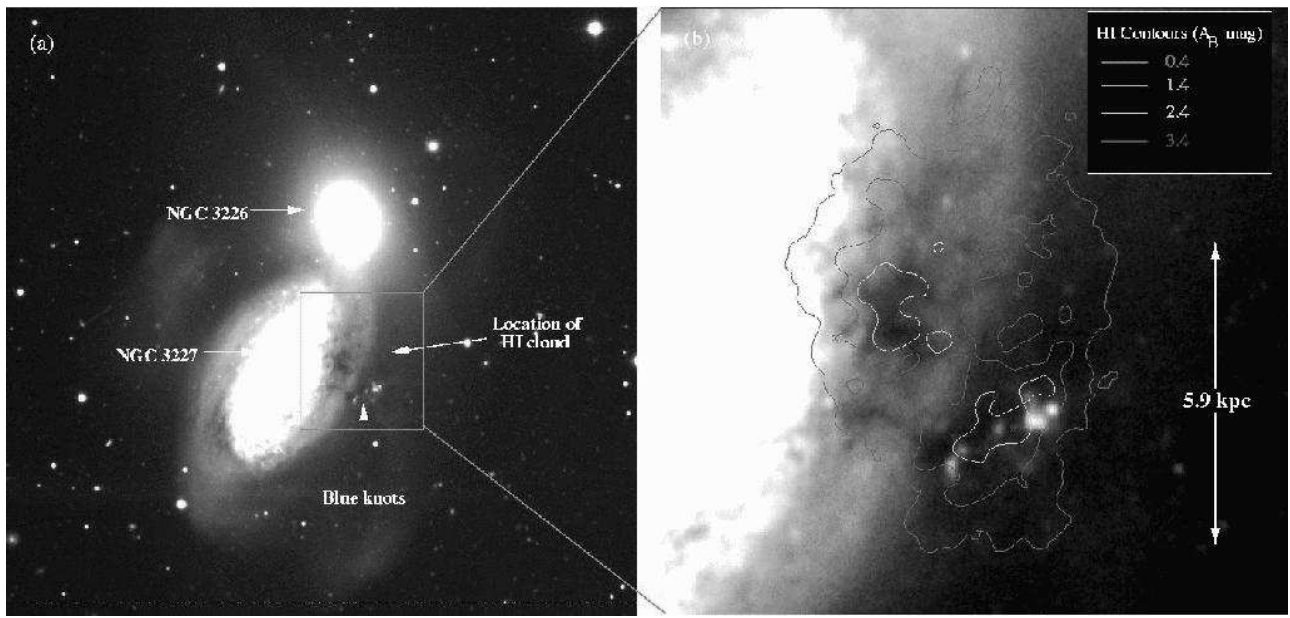} 
\caption{(a) INT $B$-band image; the location of \hi\ cloud J1023+1952 and
string of newly-discovered blue knots are indicated. (b) contours of
\hi\ emission from J1023+1952 overlaid on $B$-band image showing a close
correspondence between the high column-density \hi\ ridges and the
string of blue knots and dust patch. Contours correspond to \nh\
converted to magnitudes of optical B-band extinction assuming a
Galactic extinction law (Stavely-Smith \& Davies 1987).
\label{HIonB}} 
\end{sidewaysfigure}

The blue knots and dust structures are prominent in the B-band image
(Figure \ref{HIonB}a); as can be seen in Figure \ref{HIonB}b, the dust
is co-spatial with the \hi\ cloud J1023+1952 and, in particular, the
structure of the \hi\ northern ridge closely matches that of the dust
seen in the B-band image.  In the southern half of J1023+1952, the
string of blue knots coincides with the southern \hi\ ridge.

\begin{figure} 
\plotone{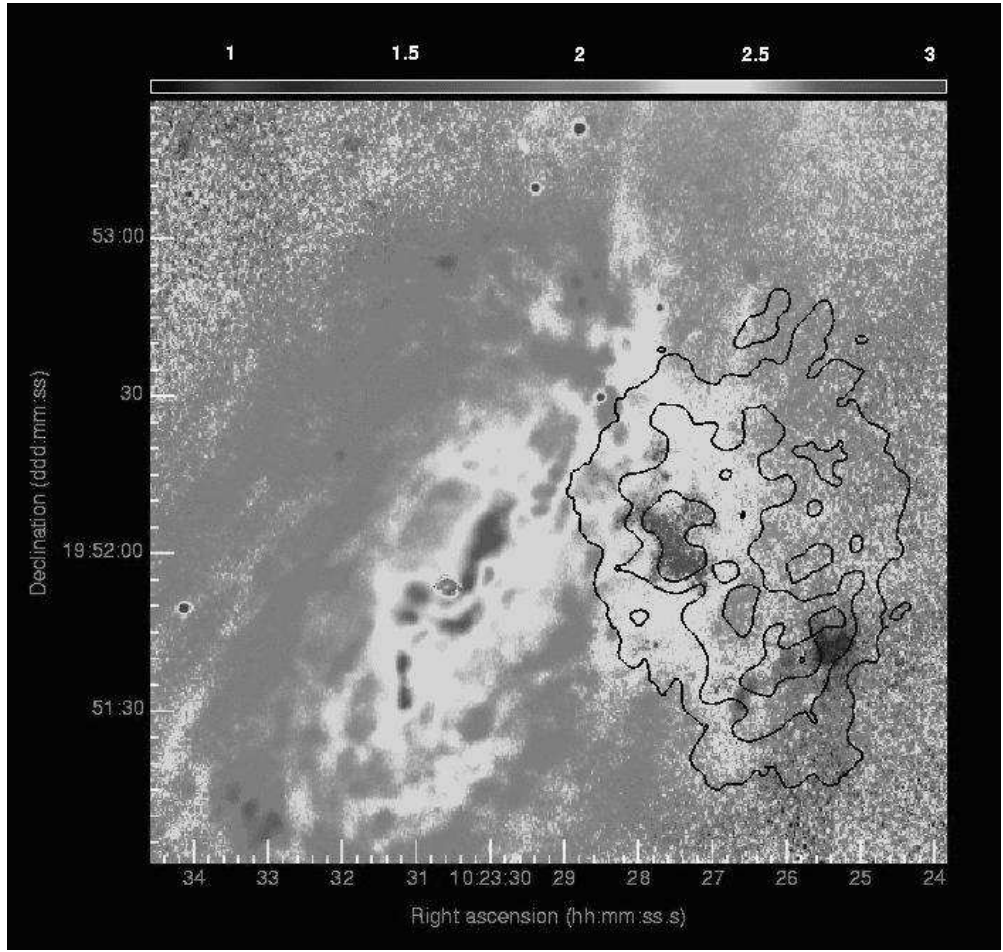} 
\caption{INT B$-$I image, corrected for Galactic extinction, with \hi\
contours overlaid. The color scale in magnitudes is indicated at the top of the
image; no correction for intrinsic extinction has been applied. The \hi\ contour levels correspond to A$_B$ = 0.4,
1.4, 2.4, 3.4 mag (as in Figure \ref{HIonB}b) .  
\label{b-i}} 
\end{figure}

The B$-$I color map in Figure \ref{b-i} shows the dust structure and
blue knots more clearly and, although no correction for intrinsic
extinction has been applied, illustrates that the knots are bluer than
any of the star formation regions in the disk of NGC~3227. Similarly
the reddest region in J1023+1952, coincident with the northern \hi\ ridge,
is redder than the rest of the NGC~3227 disk at this radius;
comparable reddening in NGC~3227 is only seen concentrated along the
known bar close to the nucleus. The lack of stellar continuum in the
NE of J1023+1952 and the close correspondence between \hi-inferred
values of A$_B$($\sim$ 2.4 mag) and B$-$I color ($\sim$3 mag) in this region
suggests that there is a significant quantity of dust intrinsic to J1023+1952
that is absorbing background stellar continuum emission from NGC~3227;
we therefore suggest that the northern half of J1023+1952 is located in
front of NGC~3227 along the line of sight. Numerical simulations of
this interacting system could help to identify projection
effects but are beyond the scope of this paper. 

\begin{figure} 
\plotone{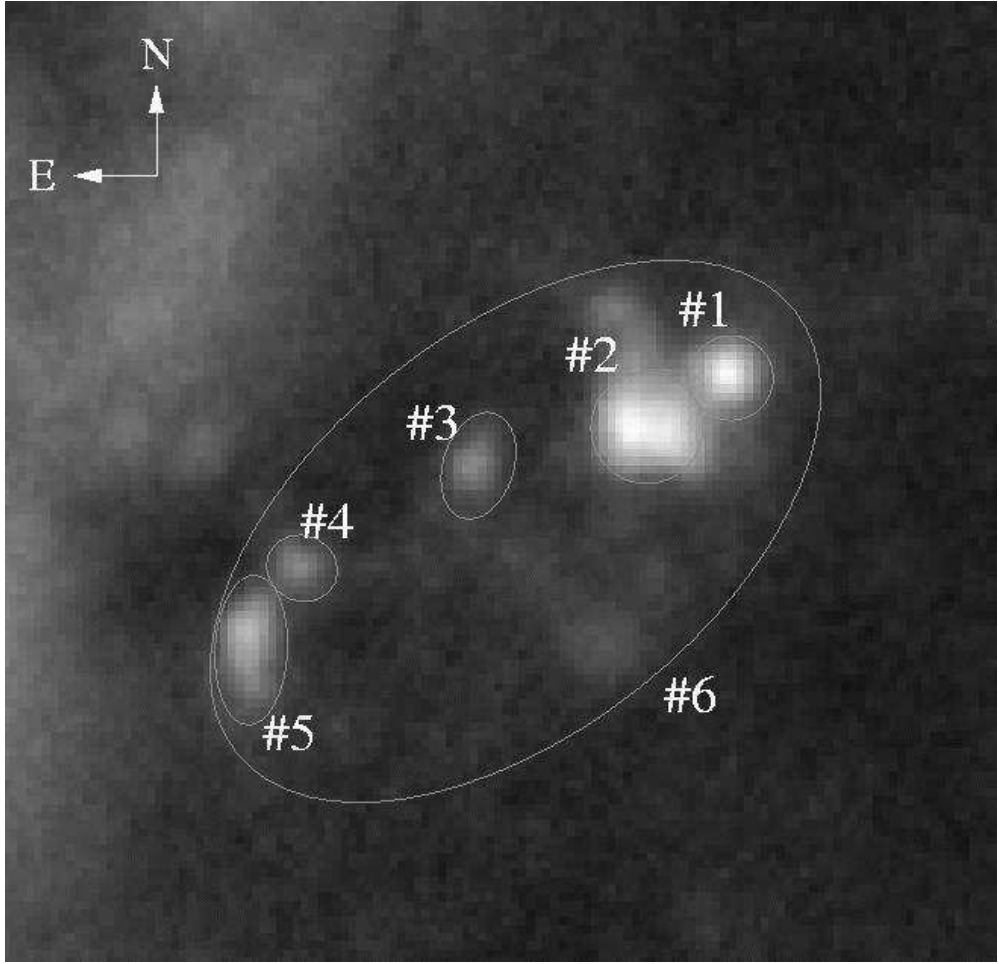} 
\caption{INT $B$-band image: apertures used to measured photometry of
knots in BRI images as listed in Table \ref{phot}. Aperture 6 has a
major axis extent of 29\arcsec.
\label{knotphot}}
\end{figure}

Table \ref{phot} lists the optical photometric
 parameters of the knots measured from the B, R and I images; the
 measurement apertures, listed in Columns (1) and (2), are shown in Figure
 \ref{knotphot}. Columns (2)$-$(4) list the measured B, R, and I magnitudes in
 the given apertures; column (4) gives the (B$-$I) color
 corrected for Galactic extinction only and column (5) lists the
 corresponding M$_B$
 calculated using a distance modulus $|~m-M~|$~=~31.57 (Tully 1988) and
 including correction for Galactic extinction. The tabulated values of
 M$_B$  represent lower limits as the intrinsic extinction is not known but
 is likely to be high. As can be seen from Figure \ref{HIonB}b,
 although the depth of the star-formation knots in the cloud along the
 line of sight is not
 known, the average intrinsic extinction could be as high as
 A$_B$~$\sim$2~mag, resulting in a total M$_B$$\sim-$15.5~mag, typical of tidal dwarf
 galaxies (e.g. Duc et al. 2000; de Oliveira et
 al. 2001). Column (7) shows the (B$-$I)$_0$ color corrected for
 Galactic extinction and additional intrinsic extinction corresponding
 to A$_B$~=~2 mag.

\subsection{Ultraviolet Continuum Emission}

\begin{sidewaysfigure} 
\plotone{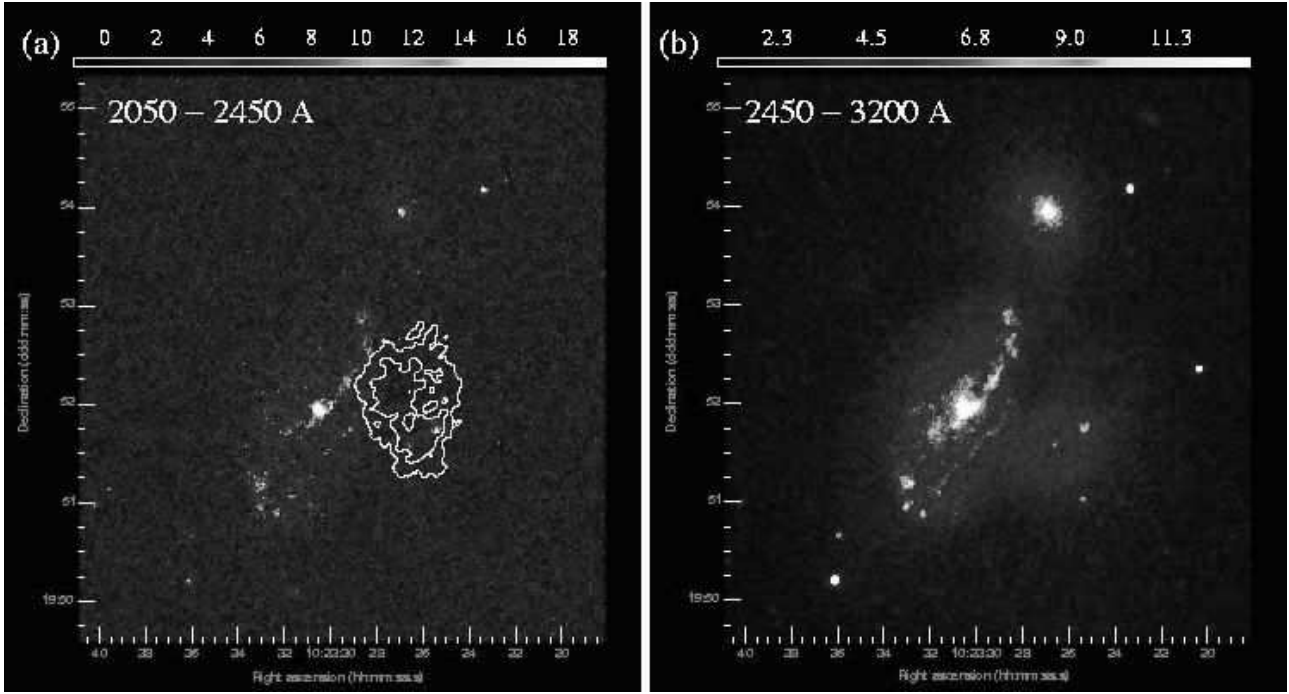} 
\caption{XMM-Newton ultraviolet images of the
NGC~3227 system in the (a) UVM2 filter with \hi\ contours overlaid (b) UVW1 filter. Flux density
scales are given in units of (a) ${\rm 10^{-16}~erg~s^{-1}~\AA^{-1}}$~\cmsq\ and (b) ${\rm 10^{-15}~erg~s^{-1}~\AA^{-1}}$~\cmsq. 
\label{uvhigar}} 
\end{sidewaysfigure}

Figure \ref{uvhigar} shows XMM-Newton ultraviolet images in the UVM2
and UVW1 filters of the NGC~3227 system. The blue
knots in J1023+1952 are still prominent at ultraviolet wavelengths
and, as can be seen in Figure \ref{uvhigar}a, are surrounded by a halo
of diffuse UV emission that is coincident with the southern half of
 J1023+1952; comparable diffuse UV emission is markedly absent from
the northern half of J1023+1952. The knots are also clearly detected at shorter
wavelengths in the less-sensitive UVM2 image (Figure
\ref{uvhigar}b). The total luminosity L$_{UV}$~(erg~s$^{-1}$~\AA$^{-1}$)
of J1023+1952 in each filter is listed in Table \ref{props}. The \hi\
distribution and \nh\ in the northern and southern \hi\ ridges in
J1023+1952 are similar, leading to the conclusion that the absence of
UV continuum emission from the northern half of J1023+1952 is due to
an intrinsic lack of star formation in this region, rather than
obscuration effects, and therefore vigorous star formation is isolated
to the southern half of J1023+1952. 
 
\subsection{H$\alpha$ Emission and Kinematics} 
\label{sectha} 

\begin{figure} 
\plotone{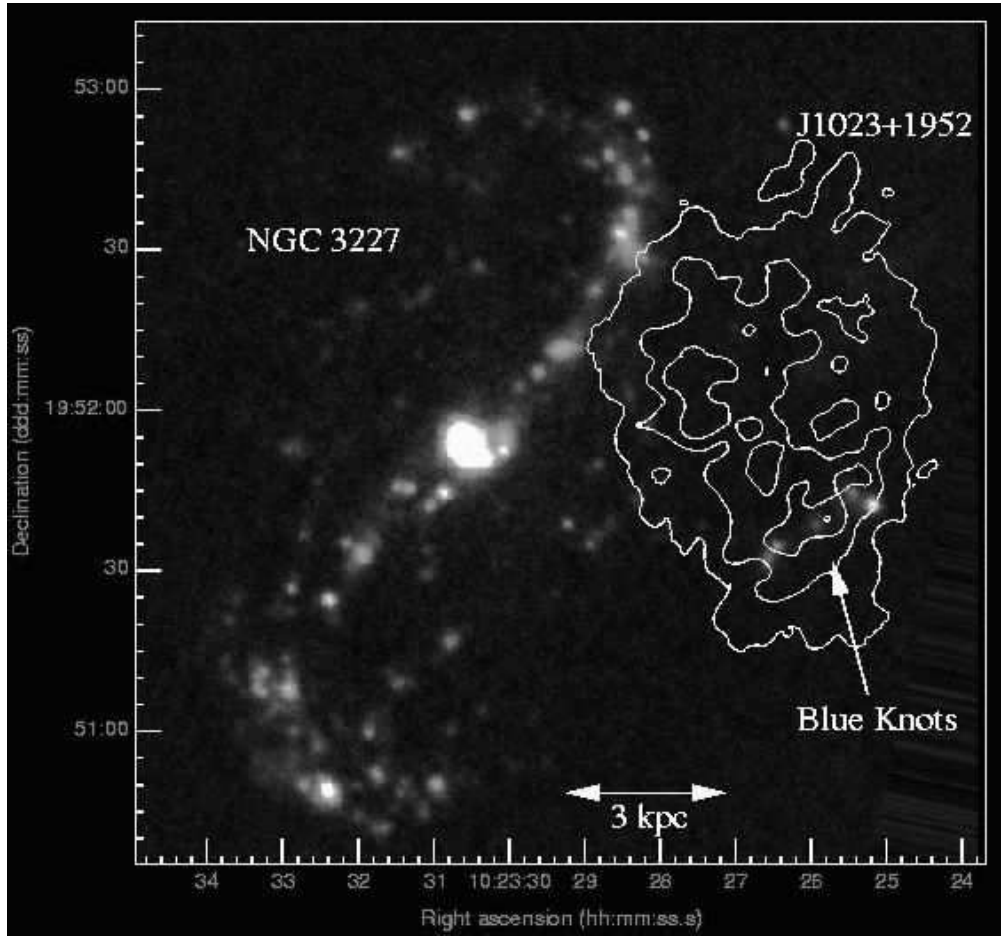} 
\caption{WHT narrow-band \ha\ image of the NGC~3227 system; \hi\
contours of J1023+1952 are overlaid.
\label{hahi}} 
\end{figure}

The presence of hot young stars in the south of J1023+1952 is further
implied by the detection of \ha\ emission from the blue
continuum-emitting knots. Figure \ref{hahi} shows the narrow-band
H$\alpha$ image of the system in which \ha\ emission from the active
nucleus and from star-forming regions in the disk and bar of NGC~3227
is evident; \ha\ emission coincident with the blue knots is
indicated. The total measured \ha\ flux is
2.55~$\times$~10$^{-14}$~erg~s$^{-1}$~cm$^{-2}$ corresponding to a
star-formation rate (after Kennicutt 1998) of
SFR~$\sim$~10.6~$\times$~10$^{-3}$~M$_{\odot}$~yr$^{-1}$ that is
uncorrected for intrinsic extinction.

\begin{figure} 
\plotone{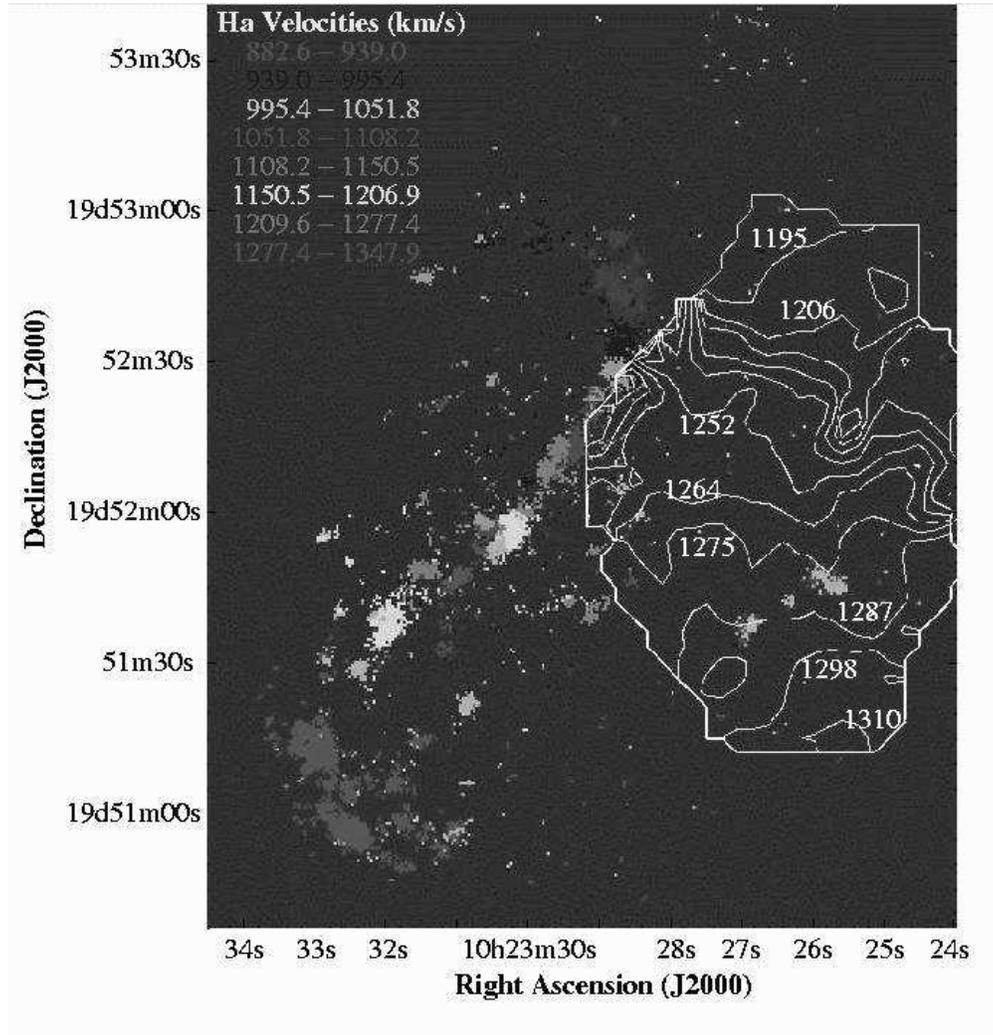} 
\caption{TAURUS \ha\ velocity field (color) of the NGC~3227 system;
\hi\ isovelocity contours for  J1023+1952 are superimposed in white and labelled in units of \kms. 
\label{hahivel}} 
\end{figure}

The ionised gas kinematics, derived from the TAURUS Fabry Perot cube,
are shown in Figure \ref{hahivel} along with the \hi\ kinematics. The
velocities of the \ha\ knots lie in the range 1210$-$1348 \kms\ and
are higher than those in the disk of NGC~3227 suggesting they are
kinematically distinct.  The corresponding \hi\ velocities measured
from the \hi\ line profile at the location of the \ha\ knots are
1290$\pm$46 \kms, closely matching those of the \ha. This strongly
suggests that the star-forming knots are embedded in J1023+1952 and
are not  merely optical knots in the background disk of NGC~3227.

\subsection{Infrared Continuum Emission}
\label{ir}

No infrared emission was detected coincident with J1023+1952.
Infrared emission was detected, as expected, from the edge of the disk
of NGC~3227 providing a consistency check on the telescope pointing
position. The lack of IR emission implies that the star formation
knots are too young and blue to produce detectable IR emission and
there is no prominent old stellar population.  The 1-$\sigma$ limiting
surface brightness sensitivity for a 1-hour exposure at K-band is 22.8
mag~arcsec$^{2}$; integrating across the \hi\ area of J1023+1952 and
using relations in Campins, Rieke \& Lebofsky (1985) and Thronson \&
Greenhouse(1988) for absolute K-band calibration and near infrared
mass-to-light ratios in galaxies, we infer an upper limit to the old
stellar population of M$_{old}<9.4\times10^{8}$~M$_{\odot}$ for
J1023+1952.

\section{DISCUSSION}

The results presented in this paper and the \hi\ kinematics presented
by Mundell et al. (1995) demonstrate that J1023+1952 is physically and
 kinematically distinct from the disk of NGC~3227 and lies in the
 foreground between us and the background disk of NGC~3227.  The close
 correspondence between the \ha\ and \hi\ velocities
 provides strong evidence that the star-forming knots are embedded in
 J1023+1952 and are not merely optical knots in the background disk of
 NGC~3227, confirming the suggestion of Mundell et al. (1995) that
 this \hi\ cloud is a dwarf galaxy. Here we discuss the possible origin of
 this dwarf galaxy as either a third, pre-existing dwarf galaxy
 involved in the interaction with NGC~3227 and NGC~3226, or a
 newly-forming dwarf galaxy condensing out of the tidal debris removed
 from the gaseous disk of NGC~3227.  Whatever the origin of
 J1023+1952, we suggest that we are witnessing shock-triggered star
 formation in infalling tidal material and that this may provide a
 mechanism for the formation of young extragalactic/intracluster HII
 regions.

\subsection{J1023+1952 $-$ Pre-existing Dwarf or Young Tidal Dwarf Galaxy?}

One scenario for the origin of J1023+1952 is that it is a pre-existing
dwarf galaxy involved in the interaction with NGC~3227 instead of, or
in addition to NGC~3226; as suggested by Mundell et al. (1995), in
this scenario the tidal forces in a prograde encounter would have
caused the northern tail to be extracted from J1023+1952 and the
southern tail to be extracted from NGC~3227, in a manner similar to
other known encounters in which two gas-rich galaxies interact
and each galaxy develops one tidal tail (e.g. Toomre \& Toomre
 1972). NGC~3227 is the brightest member of a galaxy group variously
 classified to contain between four and thirteen members (Huchra \&
 Geller 1982; Garcia 1993; Ramella, Pisani \& Geller 1997); in this
 wider group environment in which galaxy evolution is ongoing, it is
 not unreasonable to expect to find pre-existing dwarf galaxies
 (Zabludoff \& Mulchaey 2000). 

Alternatively, J1023+1952 may be newly formed from tidal debris
stripped from the disk of NGC~3227; the presence of blue optical knots
along tidal tails in interacting galaxy systems has long been
recognised as an indication that condensations of tidal debris,
ejected during an interaction, might evolve into dwarf galaxies
(e.g. Zwicky 1956; Schweizer 1978; Mirabel, Dottori \& Lutz
1992). Indeed, a chain of star-formation knots, with similar optical
properties to J1023+1952, is observed in the TDG at the end of the
southern tidal tail in NGC4038/39 (Mirabel et al. 1992).

The blue colors, large \ha\ equivalent widths and the presence of
ultraviolet emission in J1023+1952 suggests star formation is
dominated by hot young stars, in particular late-O/early-B stars with
lifetimes $<$~10$^8$ yr. Comparison of our measured UV luminosities
with those predicted by Starburst99 models (Leitherer et al. 1999) for
an instantaeous burst of star formation (including nebular continuum),
suggests a star-burst age $<$10~Myr for the knots in J1023+1952; the
high \ha\ equivalent width of $\sim$283\AA\ is also consistent with a
strong burst of star formation less than 10 Myr ago (Alonso-Herrero et
al. 1996). Mendes de Oliveira et al. (2001) used \ha\ and blue luminosities
to derive star formation rates for tidal dwarf galaxy candidates in
Stephan's Quintet. The \ha\ and blue luminosities of J1023+1952,
uncorrected for intrinsic extinction (see Table \ref{props}), imply
a star-formation rate of
SFR(\ha)~$\moresim$~10.6~$\times$10$^{-3}$ and
 SFR(L$_B$)~$\moresim$~1.1~$\times$~10$^{-3}$~M$_{\odot}$~yr$^{-1}$
 similar to giant HII regions (Mayya 1994; Figure 6 in Mendes de Oliveira et
 al 2001); correcting for extinction results in star formation rates
 similar to those of both blue compact galaxies (BCDs) (Sage et al.  1992)
 and tidal dwarf candidates in Stephan's Quintet (Mendes de Oliveira et al
 2001).  Although the \hi\ mass and \hi\ linewidth measured for
 J1023+1952 (see Table \ref{props}) are similar to
 those measured for some BCDs (van Zee, Skillman \& Salzer 1998), the
 8.9-kpc \hi\  diameter of  J1023+1952 is larger than those typical of BCD 
 ($\lesssim$2 kpc); the distribution of \hi\ and star formation is also less
 centrally concentrated in J1023+1952 than in BCDs. Tidal effects might
 explain these discrepanicies, but a striking property of J1023+1952 is
 its apparent lack of an old stellar population (Section \ref{ir}), making it an unusual
 dwarf galaxy if pre-existing (Cair\'os et al. 2003).  In the context
 of tidal dwarf galaxies, the observed \hi\ mass, column density,
 extent and global kinematics of J1023+1952
 (Tables \ref{phot} and \ref{props}) are broadly consistent with the
 range of properties known for other confirmed tidal dwarf galaxies
 (Mirabel et al. 1992; Duc \& Mirabel 1994; Hibbard et al. 1994; Duc
 et al.  2000; Braine et al. 2001; Mendes de Oliveira et al. 2001) and the
 lack of an old stellar population could be understood if J1023+1952 is
 newly formed from gaseous tidal debris removed from NGC~3227.

A key difference between classical dwarfs and TDGs is
metallicity. Unlike classical dwarf galaxies which follow a
luminosity-metallicity relation, the metallicities of TDGs appear to
lie in a narrow range 12+log(O/H)~=~8.3$-$8.7 (Duc et al. 2000;
Lisenfeld et al. 2002; Duc, Bournaud \& Masset 2004) reflecting their
origin as enriched material tidally stripped from a progenitor galaxy
disk. For the absolute magnitude of M$_B$ = -13.5, the galaxy
metallicity-luminosity (Z-L) relation (e.g. Brodie \& Huchra 1991)
would predict a metallicity of [Fe/H] $\sim$ -2 for J1023+1952. If,
however, the gas in J1023+1952 came from the disk of NGC~3227, it
would be enriched to approximately solar metallicity and hence deviate
from the classic Z-L relation, following other TDGs (Weilbacher et
al. 2003). The location of J1023+1952 on the Z-L relation will
therefore disciminate between J1023+1952 as a pre-existing or
newly-formed dwarf galaxy.

\subsection{Dark matter in J1023+1952?} 

Simulations of interacting disk galaxies with dark halos
(Barnes \& Hernquist 1992) produced tidal concentrations with less
than 5\% of their mass in dark matter, predicting that tidal dwarf
galaxies, in contrast to classical dwarf galaxies, should have low
mass-to-light ratios. Braine et al. (2001) confirmed this prediction
with detection of a significant quantity of CO in eight gas-rich
TDGs. The \hi\ distribution and kinematics of J1023+1952 are
suggestive of a rotating disk structure, but the contribution of
projection effects on the observed structure cannot be ruled
out. Assuming the neutral gas is distributed in a circularly-rotating
disk of radius $\sim$4.45~kpc the observed velocity gradient implies a
dynamical mass of
M$_D$~$\sim$3.6~$\times$~10$^9$(sin$^{-2}~i)$~M$_{\odot}$, suggesting
that \hi\ accounts for $\sim$11\% of the total mass. Correcting the
measured \hi\ mass for the presence of He such that M$_{\rm
gas}$~=~1.34~*~M$_H$ (after Hunter et al. 2000) and deriving the mass
in stars M$_{\rm star}$~=~1.54~*~L$_B$ (assuming intrinsic extinction
A$_B$=2), we conclude that the mass in gas and stars (M$_{\rm
gas}$~+~M$_{\rm star}$) accounts for at least 24\% of the dynamical
mass of J1023+1952; inclusion of the old stellar population mass upper
limit of $9.4\times10^8~M_{\odot}$ takes this value to
$\sim$50\%. Therefore, if J1023+1952 is gravitationally bound, the measured
dynamical mass would require a detection of a significant quantity of
molecular gas to exclude the presence of dark matter, given the lack
of any obvious old stellar component. Alternatively, as suggested by
the young star-formation age, J1023+1952 might not yet be fully
self-gravitating and the large \hi\ linewidth could then be explained
by a velocity gradient of infalling tidal gas rather than
rotation. Measurement of the molecular gas mass, distribution and
kinematics will provide useful constraints on the dynamics of
J1023+1952 and its current evolutionary state.

\subsection{A Dynamical Trigger for the Star Formation in J1023+1952}

There appears to be an empirical relationship between observed \hi\
column densities and the presence of star formation in galaxies
(e.g. Duc et al. 2000). The star-formation threshold is \nh\
$\sim$~0.5~$-$~1~$\times$~10$^{21}$~\cmsq\ in dwarf galaxies and low
surface brightness galaxies (e.g. Gallagher \& Hunter 1984; Skillman
1987; van der Hulst et al. 1993; van Zee et al. 1997) but Martin \&
Kennicutt (2001) find that a high gas surface density alone is not
sufficient to predict the presence of star formation in a galaxy disk
and suggest that an additional dynamical trigger is required for the
onset of star formation. Elmegreen et al. (1993) argued that large
velocity dispersions in the ISM, induced by tidal agitation, can
result in the formation of massive gas clouds similar to J1023+1952,
with high internal dispersion and high star formation efficiency.  The
velocity dispersion of \hi\ in J1023+1952 of $\sigma$~$\sim$25$-$30~\kms\
is consistent with that found by Elmegreen et al. (1993) to result in
the formation of large clouds with masses in excess of
10$^8$~M$_{\odot}$.  Barnes (2004) also argues that density-dependent
star formation rules are insufficient and suggests a dynamical trigger
in which star formation is governed by the local rate of energy
dissipation in shocks.

A dynamical trigger would appear to be necessary to explain why the
star formation in J1023+1952 is localised to the
 southern half of the \hi\ cloud, given the similarity in \nh\ in
 the southern and northern \hi\ ridges.  If
 J1023+1952 is forming from infalling tidal debris, then the southern
 ridge in J1023+1952 could
 represent a shock in the returning material where gas is compressed
 to sufficiently high gas densities to trigger the observed star
 formation.  Such a shock would compress the gas in the southern ridge
 resulting in a region in which both
 molecular and atomic densities are high, with the molecular gas
 forming {\em in situ} and providing both the fuel for the star formation
 and the necessary shielding from the resultant UV radiation (Allen et
 al. 1997).  A similar region of active star formation has also been
 observed at the base of the tidal tail of NGC~4676b (e.g. de Grijs et
 at. 2003) - a  member of the interacting galaxy pair NGC~4676
 (``The Mice"). Simulations of star formation in this system (Barnes
 2004) show that star formation occurs in a shock produced as gas from
 the curved tail falls back into its original disk.  The presence of
 broad \hi\ linewidths across J1023+1952 and, in particular, broad,
 double-peaked \hi\
 and \ha\ lines along the southern \hi\
 ridge further supports a turbulent, shock scenario for the triggering
 of the observed star formation; detection of molecular gas localized
 to the southern ridge in
 J1023+1952 with similar kinematic properties to those of the \hi\ and
 \ha\ components would provide
 additional support for the shock  hypothesis.

In the context of TDG formation, the location of J1023+1952 at the
base of the northern tidal tail is unusual.  Duc et al. (2000) argue
that, if a TDG is defined as a long-lived, gravitationally-bound
object, the most probable location for them to be observed is towards
the tips of a tidal tails since TDGs formed close to the progenitor
would disappear quickly (Mirabel et al. 1992; Duc et al. 2000;
Bournaud, Duc \& Masset 2003).  Indeed, dynamical simulations of the
merging system NGC~7252 (Hibbard \& Mihos 1995) show that material at
the base of a tidal tail is more tightly bound to the progenitor than
material at the ends and falls back quickly to small pericentric
distances, with debris at greater distances falling back more slowly
to correspondingly greater pericentric distances.  The rate of return
of tidal debris scales as t$^{-5/3}$, implying a high initial rate of
return with as much as 50\% of the tidally-stripped material falling
back to the progenitor nucleus after only 175 Myr (Hibbard \& Mihos
1995). As the projected distance between NGC~3227 and J1023+1952 along
the line of sight is not known, we cannot constrain whether J1023+1952
is stable to tidal disruption and therefore likely to be
long-lived. Nevertheless Mundell et al. (1995) note that the \hi\
distribution of the disk of NGC~3227 is asymmetrical with less \hi\ on
the side close to J1023+1952, suggesting the northern tail (and hence
the material in J1023+1952) orginated from NGC~3227. Therefore, if the
northern \hi\ tidal tail represents material stripped from the disk of
NGC~3227 and J1023+1952 is forming from returning tidal debris, the
dynamical age of J1023+1952 would be younger than 100 Myr - consistent
with the derived star formation age.

Whether J1023+1952 is a pre-existing dwarf galaxy, a young TDG or a
transient galaxy-like object, the discovery of vigorous ongoing star
formation outside of the disks of interacting galaxies such as
NGC~3227, suggests that the gas dynamics in such environments are
suitable for the shock-triggering of extragalactic star formation and may
provide a mechanism for the formation of young intergalactic HII
regions similar to those discovered by Gerhard et al. (2002) and Ryan-Weber
et al. (2004).

\section{Conclusions}
\label{conclusions}

We have used multi-wavelength imaging and spectroscopy to investigate
the nature of the massive \hi\ cloud, originally identified by Mundell
et al. (1995) in the interacting Seyfert system NGC~3227/3226 and named
J1023+1952 in this paper. We find:

\begin{itemize}
\item  Two \hi\ ridges in the northern and southern halves of the cloud, with
comparable \hi\ column densities in the range
$\sim$4~$\times$~10$^{20}$~$<$~\nh\ ~$<$~4.0~$\times$~10$^{21}$~\cmsq\
(or 0.4~$<$~A$_B$~$<$~3.4 mag).

\item  A string of blue, \ha-emitting knots of star formation, coincident
with the southern \hi\ ridge, with an extent of $\sim$2.7~kpc, total
magnitude M$_B$~$\sim$$-$15.5 mag (assuming a conservative estimate
for the internal extinction of
A$_B$$\sim$2 derived from the observed \nh\ ) and an \ha\ luminosity L$_{\ha}$~$>$~1.35$\times$10$^{39}$~erg~s$^{-1}$.

\item Ultraviolet continuum emission from the knots, a halo of diffuse
UV continuum emission around the southern half of J1023+1952 and UV
luminosities and \ha\ equivalent widths consistent with a burst of
instantaneous star formation less than 10 Myr ago.

\item Closely matched ionised and neutral gas kinematics in the
southern half of J1023+1952 with 1210~$<~V_{\ha}<~$1348~\kms\  ~and
V$_{\hi}$~$\sim$1290$\pm$46~\kms, confirming unambiguously that
 the newly-discovered star formation is occurring {\em in situ} in the
 southern half of the cloud, rather than in the underlying disk of
 NGC~3227.

\item A lack of UV and optical emission from the northern half of
J1023+1952 and a close match between the inferred reddening
 and spatial distribution of dust and \hi\ in the northern ridge,
 confirming that the northern half of the cloud lacks star
 formation and is seen primarily in absorption against the background
 stellar continuum from the disk of NGC~3227.

\end{itemize}

We conclude that J1023+1952 exhibits properties broadly consistent
with dwarf galaxies, whether pre-existing or tidal in origin; the lack
of a detectable old stellar population in J1023+1952 makes a tidal
origin more likely but measurements of metallicity are required to
confirm this. We speculate that if J1952+1052 is tidal in origin, its
unusual location at the base of the northern tidal tail and the
localization of star-formation to its southern H{\sc i} ridge suggest
that it is young and forming from gaseous material tidally stripped
from the disk of NGC~3227 that is in the first stages of infall.  We
suggest the southern ridge might represent a shock that has formed in
gas as it falls back along the northern tail onto the progenitor disk,
triggering the observed star formation. Such early TDGs are thought to
be rare due to the rapid fall-back timescale of material close to the
progenitor, suggesting a dynamical age $\tau$~$<$~100~Myr for
J1023+1952, consistent with its derived star-formation age.

\acknowledgements

CGM acknowledges financial support from the Royal Society. We are
grateful to J. Barnes for allowing us to cite his paper before
publication and thank the referee for helpful suggestions that
improved the paper. The National Radio Astronomy Observatory is a
facility of the National Science Foundation operated under cooperative
agreement by Associated Universities, Inc.  This research has made use
of NASA's Astrophysics Data System Abstract Service (ADS) and the
NASA/IPAC Extragalactic Database (NED), which is operated by the Jet
Propulsion Laboratory, California Institute of Technology, under
contract with the National Aeronautics and Space Administration. The
WHT and INT are operated on the island of La Palma by the Isaac Newton
Group in the Spanish Observatorio del Roque de los Muchachos of the
Instituto de Astrofisica de Canarias. The United Kingdom Infrared
Telescope is operated by the Joint Astronomy Centre on behalf of the
UK Particle Physics and Astronomy Research Council.




\clearpage


\begin{table}
\caption{BRI aperture photometry (see Figure \ref{knotphot} for apertures)}
\label{phot}
\begin{tabular}{lccccccc}
\hline
ID & Aperture Diameter&  B &  R   &  I  &$(B-I)$& $M_B$ & $(B-I)_0$\\
  & (arcsec)&(mag)&(mag) &(mag)&(mag) & (mag)&(mag)\\
(1) &   (2) &  (3)   & (4)& (5) & (6)& (7)\\
\hline
\#1 & 3.46&20.67&19.75&19.49&	1.13&$-$11.0& 0.03\\	    	  
\#2 & 4.58&19.89&19.16&18.63&	1.21&$-$11.8& 0.11\\	    	  
\#3 & 4.49&21.30&20.13&19.53&	1.72&$-$10.4& 0.62\\	    	  
\#4 & 2.97&21.69&20.47&19.86&	1.78&$-$10.0& 0.68\\	    	  
\#5 & 6.14&20.74&19.54&18.80&	1.89&$-$10.9& 0.79\\	    	  
\#6 & 29.1$^a$&18.16&16.98&16.48&	1.63&$-$13.5& 0.53\\
\hline
&$^a$Ellipse eccentricity is 0.83\\
\end{tabular}
\end{table}

\begin{table}
\caption{Derived Properties for J1023+1952}
\label{props}
\begin{tabular}{lc}
\hline
$\alpha$ (J2000)&10h 23m 26\fsec5\\
$\delta$ (J2000)&19\arcdeg\ 52\arcmin\ 0\farcs0\\
\hi\ Diameter (kpc) & 8.9\\
M$_H$ (M$_{\odot}$)& 3.8~$\times$~10$^8$\\
Systemic Velocity (\kms)& $\sim$1270\\ 
W$_{20}$ (\kms)& 120\\
Dynamical Mass M$_D$ (M$_{\odot}$)&3.6~$\times$~10$^9$\\
Star-formation Extent (kpc)&2.7\\
$^a$Total M$_B$ (mag)&$-$13.5\\
\ha\ Equivalent Width (\AA)& 283\\
$^a$Total F$_{H\alpha}$ (erg~s$^{-1}$ \cmsq)&2.55~$\times$~10$^{-14}$\\
$^a$Total L$_{H\alpha}$ (erg~s$^{-1}$)&1.35~$\times$~10$^{39}$\\
$^a$Total L$_B$ (~L$_{\odot}$)&3.9~$\times$~10$^7$\\
$^a$Total L$_{\lambda2905~\AA}$ (erg~s$^{-1}$~\AA$^{-1}$)&5.6~$\times$~10$^{38}$\\
$^a$Total L$_{\lambda2298~\AA}$ (erg~s$^{-1}$~\AA$^{-1}$)&7.5~$\times$~10$^{36}$\\
\hline
$^a$Uncorrected for extinction
\end{tabular}
\end{table}

\clearpage				

\begin{thebibliography}{}

\bibitem[Allen97]{al97} Allen, R.J., Knapen, J.H., Bohlin, R. \&
Stecher, T.P. 1997, \apj, 487, 171

\bibitem[ah96]{ah96} Alonso-Herrero, A., Aragon-Salamanca, A., Zamorano, J. \&
Rego, M.  1996 \mnras, 278, 417  

\bibitem[Barn92]{bh92} Barnes, J. \& Hernquist, L. 1992 ARA\&A, 30,
705

\bibitem[Barnes04]{b04}Barnes, J.E 2004 \mnras, 350, 798

\bibitem[bourn03]{bourn03} Bournaud, F, Duc, P.-A. \& Masset, F. 2003
A\&A, 411, L469

\bibitem[Braine01]{br01} Braine, J.,  Duc, P.-A., Lisenfeld, U.,
Charmandaris, V., Vallejo, O., Leon, S. \& Brinks, E. 2001 A\&A, 378, 51

\bibitem[brh91]{brh91} Brodie J. \& Huchra J. 1991, \apj, 379, 157

\bibitem[Burkey94]{bur94} Burkey, J.M., Keel, W.C., Windhorst, R.A. \&
Franklin, B.E. 1994, \apj, 429, L13

\bibitem[crl85]{crl85} Campins, H., Rieke, G.H. \& Lebofsky, M.J. 1985 \aj,
90, 896

\bibitem[cair03]{cair03}Cair\'os, L.M.; Caon, N., 
Papaderos, P., Noeske, K., V\'ilchez, J.M.,
Lorenzo, B.G. \& Mu\~noz-Tu\~n\'on, C. 2003 \apj, 593, 312 

\bibitem[Cowie96]{co96} Cowie, L.L., Songaila, A., Hu, E.M. \& Cohen,
J.G. 1996, \aj, 112, 839

\bibitem[deGr03]{deG03} de Grijs, R., Lee, J.T., Clemencia Mora Herrera, M.,
Fritze- v. Alvensleben, U. \& Anders, P. 2003, New Astr. 8, 155 

\bibitem[Duc94]{duc94} Duc, P.-A. \& Mirabel, I.F. 1994, A\&A, 289, 83

\bibitem[Duc00]{duc00} Duc, P.-A., Brinks, E., Springel, V., Pichardo,
B., Weilbacher, P. \& Mirabel, I.F. 2000, \aj, 120, 1238

\bibitem[Duc04]{Duc04} Duc, P.-A., Bournaud, F. \& Masset, F. 2004, in
IAU Symposium 217, "Recycling intergalactic and interstellar matter",
eds. Duc, P.~A., Braine, J.,  Brinks, E., ASP,  in press (astro--ph/0402252)

\bibitem[Elme93]{elm93} Elmegreen, D.M.,  Kaufman, M. \& Thomasson,
M. 1993, \apj, 412, 90

\bibitem[gh84]{gh4} Gallagher, J.S. \& Hunter, D.A. 1984, \araa, 22 37

\bibitem[gar93]{gar93} Garcia, A.M. 1993 A\&AS, 100, 47


\bibitem[ger02]{g02} Gerhard, O., Anaboldi, M., Freeman, K.C. \&
Okamura, S. 2002, \apj, 580, L121

\bibitem[hib94]{hib94} Hibbard, J.E., Guhathakurta, P., van Gorkom,
J.H. \& Schweizer, F. 1994 \aj, 107, 67 

\bibitem[hib95]{hib95} Hibbard, J.E. \& Mihos, J.C. 1995 \aj, 110, 140

\bibitem[hg82]{hg82} Huchra, J.P. \& Geller, M.J. 1982 \apj, 257, 423

\bibitem[Huns96]{huns96} Hunsberger, S.D., Charlton, J.C. \& Zaritsky,
D. 1996, \apj, 462, 50

\bibitem[Hunter00]{hun00} Hunter, D.A., Hunsberger, S.D. \& Roye,
E.W. 2000, \apj, 542, 137

\bibitem[James04]{jam04} James, P.A., Shane, N.S., Beckman, J.E. et al. 2004, A\&A, 414, 23


\bibitem[Kenn98]{K98} Kennicutt, R.C. 1998, \araa, 36, 189 


\bibitem[Leith99]{leith99} Leitherer, C., Schaerer, D., Goldader,J.D.,
Gonz\'alez-Delgado, R.M., Robert, C., Kune, D.F., de Mello,D.F.,
Devost, D. \& Heckman, T.M. 1999, \apjs, 123, 3

\bibitem[lis02]{lf02} Lisenfeld, U., Braine, J., Vallejo, O., Duc,
P.-A., Leon, S., Brinks, E. \& Charmandaris, V. 2002, in "Modes of
Star Formation and the Origin of Field Populations", eds. E.K. Grebel
\& W. Brandner, ASP Conf. Series Vol.  285, 406

\bibitem[mart01]{mk01} Martin, C.L. \& Kennicutt, R.C. 2001, \apj,
555, 301

\bibitem[Mason01]{mas01} Mason, K.O., Breeveld, A., Much, R. et al. 2001, A\&A, 365, L36

\bibitem[Mayya94]{may94} Mayya, Y.D. 1994, \aj, 108, 1276

\bibitem[McAl83]{mca83} McAlary, C.W., McLaren, R.A., McGonegal,
R.J. \& Maza, J. 1983, \apjs, 52, 341 

\bibitem[Mendes01]{men01} Mendes de Oliveira, C., Plana, H., Amram, P.,
Balkowski, C. \& Bolte, M. 2001, \aj, 121, 2524


 
\bibitem[mdl92]{mdl92} Mirabel, I.F., Dottori, H. \& Lutz, D. 1992 A\&A, 256, L19
 
\bibitem[Mihos01]{mih01} Mihos, C. 2001 ApJ, 550, 94

\bibitem[Mun95]{mun95} Mundell, C.G., Pedlar, A., Axon, D.J., Meaburn,
J. \& Unger, S.W. 1995, \mnras, 277, 641 

\bibitem[Mundell 1999]{M99} Mundell, C.G., Pedlar, A., Shone, D.L. \& Robinson,
A. 1999 \mnras, 304, 481

\bibitem[Ram97]{ram97} Ramella, M., Pisani, A. \& Geller, M.J. 1997 \aj, 113, 483

\bibitem[RyW03]{rw03} Ryan-Weber, E.V. et al. 2004 \aj, 127, 1431

\bibitem[Sage92]{sage92} Sage, L.J., Saltzer, J.J., Loose, H.-H. \&
Henkel, C. 1992, A\&A, 265, 19

\bibitem[schw78]{schw78} Schweizer, F. 1978, in "Structure and Properties of Nearby Galaxies", IAU Symp. 77, 279 

\bibitem[skill87]{sk87} Skillman, E.D. 1987, in "Star Formation in
Galaxies", ed. C.J. Lonsdale Persson, NASA, CP-2466, p. 263

\bibitem[Stave87]{stav87} Staveley-Smith, L. \& Davies, R.D. 1987,
MNRAS, 224, 953 

\bibitem[Struck99]{str99} Struck, C. 1999, Ph. R., 321, 1


\bibitem[tg88]{tg88} Thronson H.A., Jr. \& Greenhouse, M.A.  1988 ApJ,
325, 604

\bibitem[Toom72]{tom72} Toomre, A. \& Toomre, J. 1972, ApJ 178, 623

\bibitem[Tully88]{tul88} Tully, R.B. 1988, Nearby Galaxies Catalog
(Cambridge: Cambridge Univ. Press)

\bibitem[vh93]{vh93} van der Hulst, J.M., Skillman, E.D., Smith, T.R., Bothun, G.D.,
McGaugh, S.S. \& de Blok, W.J.G. 1993, \aj, 106, 548

\bibitem[vanz97]{vz97} van Zee, L., Haynes, M.P., Salzer, J.J. \&
Broeils, A.H. 1997, \aj, 113, 1618

\bibitem[vanz98]{vz98} van Zee, L., Skillman, E.D. \& Salzer,
J.J. 1998 \aj, 116, 1186

\bibitem[weil03]{weil03} Weilbacher, P.M., Duc, P.-A.,
Fritze-v. Alvensleben, U. 2003, A\&A, 397, 545

\bibitem[Zab00]{zm00} Zabludoff, A.I. \& Mulchaey, J.S. 2000, \apj, 539, 136

\bibitem[Zwicky56]{zw56} Zwicky, F. 1956, Erge. Exacten Naturwiss.,
29, 344 


\end{thebibliography}
\end{document}